\documentclass[onecolumn,showpacs,floatfix,aps,nofootinbib]{revtex4}
\usepackage{amssymb,amsmath}
\usepackage[dvips]{graphics,color}
\usepackage{epsfig}\usepackage{float}
\newcommand{\ba}{\begin{eqnarray}}
\newcommand{\ea}{\end{eqnarray}}
\newcommand{\bege}{\begin{equation}}
\newcommand{\enge}{\end{equation}}
\newcommand{\beq}{\begin{eqnarray}}
\newcommand{\benu}{\begin{enumerate}}
\newcommand{\enu}{\end{enumerate}}
\newcommand{\eeq}{\end{eqnarray}}

\newcommand{\noi}{\noindent}

\newcommand{\ka}{\kappa}
\newcommand{\mr}{\mathring}
\newcommand{\n}{\nonumber\\}
\newcommand{\g}{\gamma}
\newcommand{\RR}{\mathbb{R}}
\newcommand{\mt}{\mathcal}
\newcommand{\cl}{\mt{C}\ell}

\newcommand{\w}{\wedge}
\newcommand{\vv}{{\bf v}}
\newcommand{\uu}{{\bf u}}
\newcommand{\ww}{{\bf w}}
\newcommand{\CC}{\mathbb{C}}
\newcommand{\OO}{\mathbb{O}}
\newcommand{\II}{\mathbb{I}}
\newcommand{\mmu}{\mathfrak{u}}
\newcommand{\me}{\frac{1}{2}}
\begin{document}

\title{From Dirac spinor fields to ELKO}

\author{R. da Rocha}
\email{roldao.rocha@ufabc.edu.br, roldao@ifi.unicamp.br} \affiliation{
Centro de Matem\'atica, Computa\c c\~ao e Cogni\c c\~ao,
Universidade Federal do ABC, 09210-170, Santo Andr\'e, SP, Brazil\\and\\
Instituto de F\'{\i}sica ``Gleb Wataghin''\\ Universidade
Estadual de Campinas, Unicamp\\
 13083-970 Campinas, S\~ao Paulo, Brasil}
\author{J. M. Hoff da Silva}
\email{hoff@ift.unesp.br} \affiliation{Instituto de F\'{\i}sica
Te\'orica, Universidade Estadual Paulista, Rua Pamplona 145
01405-900 S\~ao Paulo, SP, Brazil}

\begin{abstract}
Dual-helicity eigenspinors of the charge conjugation
operator (ELKO spinor fields) belong --- together with Majorana spinor fields --- to a wider class of spinor fields, 
the so-called flagpole spinor fields, corresponding to the class (5), according to Lounesto spinor field classification based
on the relations and values taken by their associated bilinear covariants. There exists only six such disjoint classes: the first three 
corresponding to Dirac spinor fields, 
and the other three respectively corresponding to flagpole, flag-dipole and Weyl spinor fields. 
This paper is devoted to investigate and provide the necessary and sufficient conditions to map 
Dirac spinor fields to ELKO, in order to naturally extend the Standard Model to spinor fields possessing mass dimension one.
As ELKO is a prime candidate to describe dark matter, an adequate and necessary formalism is introduced and developed here, 
to better understand the algebraic, geometric and physical properties of ELKO spinor fields, and their underlying relationship 
to Dirac spinor fields.
 
\end{abstract}
\maketitle
\section{Introduction}

ELKO --- \emph{Eigenspinoren des Ladungskonjugationsoperators} --- spinor
fields\footnote{ELKO is the German acronym for Dual-helicity eigenspinors of the charge conjugation
operator \cite{allu}.}  represent an extended set of
Majorana spinor fields,  describing a  non-standard Wigner class of fermions, in which the charge conjugation
and the parity operators commute, rather than anticommute \cite{allu,alu2,alu4}. Further, ELKO  
accomplishes dual-helicity eigenspinors of the spin-1/2 charge conjugation operator, 
and carry mass dimension one, besides having non-local properties.
In order to find an adequate mathematical formalism for representing dark
matter by a spinor field associated with mass dimension one,
 Ahluwalia-Khalilova and Grumiller have just ushered the
ELKO \cite{allu} into quantum field theory, and it has also given rise to subsequent applications in cosmology. 
ELKO is a representative of a
neutral fermion 
described by a set of four spinor fields, 
two of which are identified to massive McLennan-Case (Majorana) spinor fields \cite{mac,case},  and other two which were not known yet. 
Another surprising character involving ELKO is that its
 Lagrangian possesses interaction  neither with Standard Model fields nor with gauge fields, which endows ELKO to  
be a prime candidate to describe dark matter \cite{alu3,ani,123}, which has recent observational confirmation \cite{jee}. Likewise,
 the Higgs boson can interact with ELKO, and it also could be tested at LHC.

In
the low-energy limit, ELKO behaves as a representation of the Lorentz group.
However, all spinor fields in Minkowski spacetime can be given --- from the 
classical viewpoint\footnote{It is well known that spinors have three different,
although equivalent, definitions:
the operatorial, the classical and the algebraic one \cite{moro,rod,op,hes,hes1,cru,benn,cartan,chev,riesz}.} --- as
 elements of the carrier spaces of the $D^{(1/2,0)}\oplus D^{(0,1/2)}$
or $D^{(1/2,0)}$, or $D^{(0,1/2)}$ representations of SL(2,$\mathbb{C)}$. 
P. Lounesto, in the
classification of spinor fields, proved that any spinor field 
belongs to one of the six classes found by him \cite{lou1,lou2}. Such an \textit{algebraic} classification is
based on the values assumed by their bilinear covariants, the Fierz
identities, aggregates and boomerangs 
\cite{lou1,lou2,meu}. Lounesto spinor field classification has wide applications in cosmology and astrophysics (via ELKO, for instance see 
\cite{allu,alu2,alu3,boe1,boe2,meu}), and 
in General Relativity: it was recently demonstrated that Einstein-Hilbert, the Einstein-Palatini, and the 
Holst\footnote{The Holst action is shown to be equivalent to the Ashtekar
 formulation of Quantum Gravity \cite{hor}.} actions
 can be derived from the Quadratic Spinor Lagrangian (that describes supegravity) \cite{tn1,bars1}, 
when the three classes of Dirac 
spinor fields, under Lounesto spinor field classification, are considered \cite{jpe}. 
It was also shown \cite{meu} that ELKO represents a larger class of Majorana spinor fields,
and that those spinor fields covers one of the six classes in Lounesto spinor field classification.
ELKO possesses an intrinsic and genuine geometric structure
behind, and a great variety of geometrical and algebraic concepts, and their applications in Physics and 
Mathematical-Physics, 
e.g., the formalism of Penrose twistors, flagpoles and flag-dipoles \cite{pe1,pe2,pe3,pos1,rocha1,dor2,bette,bette2}, 
can be unified, described, and generalized via this formalism.

One of the main purposes of this paper 
is to analyze and investigate the underlying equivalence between Dirac spinor fields (DSFs) and ELKO, i.e., 
under which conditions a DSF can be led 
to an ELKO, since they are inherently distinct and represent disjoint classes in Lounesto spinor field classification. 
For instance, while the latter belongs to class (5) 
 under such classification, the former is a representative of spinor fields of  
types-(1), -(2), and -(3). In addition, when acting on ELKO, the parity $\mathbb{P}$ and charge conjugation $C$ 
operators \emph{commutes} and $\mathbb{P}^2 = -1$, 
while when acting upon Dirac spinor fields, such operators \emph{anticommutes} and $\mathbb{P}^2 = 1$.
Besides, $C\mathbb{P}\mathbb{T}$ equals $+1$ and $-1$, respectively for DSFs and ELKO.
Any invertible map that takes Dirac
particles and leads to ELKO is also capable to make
mass dimension transmutations, since DSFs present mass dimension three-halves, instead of mass dimension one
associated with ELKO. The main physical motivation of this 
paper\footnote{R. da Rocha thanks to Prof. Dharamvir Ahluwalia-Khalilova for 
 private communication on the subject.} is to provide the initial pre-requisites to construct
 a natural extension of the Standard
Model (SM) in order to incorporate ELKO, and consequently a possible description of dark matter \cite{allu,alu2,alu3} in this context.

The paper is organized as follows: after briefly  presenting some essential algebraic preliminaries in Section (II), we  introduce
 in 
Section (III) the bilinear covariants
together with the Fierz
identities. Also, the Lounesto
classification of spinor fields is presented together with the definition
of ELKO spinor fields \cite{allu}, showing that ELKO is indeed a flagpole
spinor field with opposite (dual) helicities \cite{allu,alu2,meu}. In Section (IV) the mapping from Dirac spinor fields to ELKO is widely investigated in details.

\section{Preliminaries}
\label{w2}

Let $V$ be a finite $n$-dimensional real vector space and $V^*$
denotes its dual. We consider the tensor algebra
$\oplus_{i=0}^\infty T^i(V)$ from which we restrict our
attention to the space $\Lambda(V) =
\oplus_{k=0}^n\Lambda^k(V)$ of multivectors over $V$.
$\Lambda^k(V)$ denotes the space of the antisymmetric
 $k$-tensors, isomorphic to the  $k$-forms vector space.  Given
$\psi\in\Lambda(V)$, $\tilde\psi$ denotes the \emph{reversion},
 an algebra antiautomorphism
 given by $\tilde{\psi} = (-1)^{[k/2]}\psi$ ([$k$] denotes the integer
part of $k$).
  If $V$ is endowed with a non-degenerate, symmetric, bilinear map $g: V^*\times V^* \rightarrow \RR$, it is possible to extend $g$ to
$\Lambda(V)$. Given $\psi=\uu^1\w\cdots\w \uu^k$ and
$\phi=\vv^1\w\cdots\w \vv^l$, for $\uu^i, \vv^j\in V^*$, one
defines $g(\psi,\phi)
 = \det(g(\uu^i,\vv^j))$ if $k=l$ and $g(\psi,\phi)=0$ if $k\neq l$. The
projection of a multivector $\psi= \psi_0 + \psi_1 + \cdots +
\psi_n$,
 $\psi_k \in \Lambda^k(V)$, on its $p$-vector part is given by
$\langle\psi\rangle_p$ = $\psi_p$.
 Given $\psi,\phi,\xi\in\Lambda(V)$, the  {\it left contraction} is
defined implicitly by
$g(\psi\lrcorner\phi,\xi)=g(\phi,\tilde\psi\w\xi)$.
 For $a \in \RR$, it follows that
 ${\bf v} \lrcorner a = 0$. The
 {\it right contraction} is analogously defined by
$g(\psi\llcorner\phi,\xi)=g(\phi,\psi\w\tilde\xi)$.
 Both
contractions are related by ${\bf v} \lrcorner \psi = -\hat\psi
\llcorner {\bf v}$. The Clifford product between $\ww\in V$ and
$\psi\in\Lambda(V)$ is given by $\ww\psi = \ww\w \psi + \ww\lrcorner
\psi$.
 The Grassmann algebra $(\Lambda(V),g)$
endowed with the Clifford  product is denoted by $\cl(V,g)$ or
$\cl_{p,q}$, the Clifford algebra associated with $V\simeq
\RR^{p,q},\; p + q = n$.  In what follows $\RR,\CC$  denote
respectively the real and complex numbers.

\section{Bilinear Covariants and ELKO spinor fields}

This Section is devoted to recall the bilinear covariants, using the programme introduced in 
\cite{meu}, which we briefly recall here.
In this article all spinor fields live in Minkowski spacetime
$(M,\eta ,D,\tau_{\eta},\uparrow)$. The manifold $M$
$\simeq\mathbb{R}^{4}$, $\eta$ denotes a constant metric, where $\eta(\partial/\partial x^{\mu},\partial/\partial x^{\nu
})=\eta_{\mu\nu}=\mathrm{diag}(1,-1,-1,-1)$, $D$ denotes the Levi-Civita connection associated with
$\eta$, $M$ is oriented by the 4-volume element
$\tau_{\eta}$ and
time-oriented by $\uparrow$.  Here $\{x^{\mu}\}$ denotes global
coordinates in the Einstein-Lorentz gauge, naturally
adapted to an inertial reference frame  $\mathbf{e}_{0}%
=\partial/\partial x^{0}$. Let 
$\mathbf{e}_{i}=\partial/\partial x^{i}$, $i=1,2,3$. 
 Also,
$\{\mathbf{e}_{\mu}\}$ is a section of the frame bundle
$\mathbf{P}_{\mathrm{SO}_{1,3}^{e}}(M)$ and
$\{\mathbf{e}^{\mu}\}$ is its reciprocal frame satisfying $\eta(\mathbf{e}%
^{\mu},\mathbf{e}_{\nu}):=\mathbf{e}^{\mu}\cdot\mathbf{e}_{\nu}=\delta_{\nu
}^{\mu}$. Classical spinor
fields  
carrying a $D^{(1/2,0)}\oplus D^{(0,1/2)}$,
or $D^{(1/2,0)}$, or $D^{(0,1/2)}$ representation of
SL$(2,\mathbb{C)\simeq
}\;\,\mathrm{Spin}_{1,3}^{e}$ are sections of the vector bundle 
$
\mathbf{P}_{\mathrm{Spin}_{1,3}^{e}}(M)\times_{\rho}\mathbb{C}^{4},
$
where $\rho$ stands for the $D^{(1/2,0)}\oplus D^{(0,1/2)}$ (or
$D^{(1/2,0)}$
or $D^{(0,1/2)}$) representation of SL$(2,\mathbb{C)\simeq}\;\,\mathrm{Spin}%
_{1,3}^{e}$ in $\mathbb{C}^{4}$.  Given a spinor field
$\psi$ $\in
\sec\mathbf{P}_{\mathrm{Spin}_{1,3}^{e}}(M)\times_{\rho}\mathbb{C}^{4}$
the bilinear covariants are the following sections of
 the exterior algebra bundle of \textit{multivector} fields
 \cite{moro}:
\begin{align}
\sigma &  =\psi^{\dagger}\gamma_{0}\psi,\quad\mathbf{J}=J_{\mu}\mathbf{e}%
^{\mu}=\psi^{\dagger}\gamma_{0}\gamma_{\mu}\psi\mathbf{e}^{\mu},\quad
\mathbf{S}=S_{\mu\nu}\mathbf{e}^{\mu\nu}=\frac{1}{2}\psi^{\dagger}\gamma
_{0}i\gamma_{\mu\nu}\psi\mathbf{e}^{\mu}\wedge\mathbf{e}^{\nu},\nonumber\\
\mathbf{K} &  = K_\mu\mathbf{e}^\mu = \psi^{\dagger}\gamma_{0}i\gamma_{0123}\gamma_{\mu}%
\psi\mathbf{e}^{\mu},\quad\omega=-\psi^{\dagger}\gamma_{0}\gamma_{0123}%
\psi,\label{fierz}%
\end{align}
The set $\{\gamma_{\mu}\}$ refers to
the Dirac matrices in
chiral representation (see Eq.(\ref{dirac matrices})). Also  
$
\{\mathbf{1}_{4},\gamma_{\mu},\gamma_{\mu}\gamma_{\nu},\gamma_{\mu}\gamma
_{\nu}\gamma_{\rho},\gamma_{0}\gamma_{1}\gamma_{2}\gamma_{3}\}$ ($\mu,\nu,\rho=0,1,2,3$, and $\mu<\nu<\rho$) is a
basis for $\mathbb{C}(4)$  satisfying
 \cite{lou1}
$\gamma_{\mu}\gamma_{\nu}+\gamma_{\nu}\gamma_{\mu}   =2\eta_{\mu\nu}
\mathbf{1}_{4}$ and the Clifford
product  is denoted by juxtaposition. More
details on  notations can be found in
\cite{moro,rod}.

Given a fixed spin frame the bilinear covariants are considered
as being the following \textit{operator} fields, for each $x\in M$, as mappings $\mathbb{C}^{4}%
\rightarrow\mathbb{C}^{4}$:
\begin{align}
\sigma &
=\psi^{\dagger}\gamma_{0}\psi,\quad\mathbf{J}=J_{\mu}\gamma^{\mu
}=\psi^{\dagger}\gamma_{0}\gamma_{\mu}\psi\gamma^{\mu},\quad\mathbf{S}%
=S_{\mu\nu}\gamma^{\mu\nu}=\frac{1}{2}\psi^{\dagger}\gamma_{0}i\gamma_{\mu\nu
}\psi\gamma^{\mu\nu},\nonumber\\
\quad\mathbf{K}  & =K_\mu\gamma^\mu  =\psi^{\dagger}\gamma_{0}i\gamma_{0123}\gamma_{\mu}%
\psi\gamma^{\mu},\quad\omega=-\psi^{\dagger}\gamma_{0}\gamma_{0123}\psi.
\label{11}%
\end{align}\noindent In the case of the electron, described by Dirac spinor
fields (classes 1, 2 and 3 below), $\mathbf{J}$ is a
future-oriented timelike current vector which gives the current of
probability, the bivector $\mathbf{S}$ is associated with the
distribution of intrinsic angular momentum, and the spacelike
vector $\mathbf{K}$ is associated with the direction of the
electron spin. For a detailed discussion concerning such entities,
their relationships and physical interpretation, and
generalizations, see, e.g., \cite{cra,lou1,lou2,holl,hol}.

The bilinear covariants satisfy the Fierz identities
\cite{cra,lou1,lou2,holl,hol}
\begin{equation}
\mathbf{J}^{2}=\omega^{2}+\sigma^{2},\quad\mathbf{K}^{2}=-\mathbf{J}^{2}%
,\quad\mathbf{J}\llcorner\mathbf{K}=0,\quad\mathbf{J}\wedge\mathbf{K}%
=-(\omega+\sigma\gamma_{0123})\mathbf{S}. \label{fi}%
\end{equation}
\noindent 
A spinor field such that \emph{not both} $\omega$ and $\sigma$ are
null is said to be regular. When $\omega=0=\sigma$, a spinor field
is said to be \textit{singular}. 

Lounesto spinor field classification is given by the following
spinor field classes \cite{lou1,lou2}, where in the first three
classes it is implicit that $\mathbf{J}$\textbf{,
}$\mathbf{K}$\textbf{, }$\mathbf{S}$ $\neq0$:

\begin{itemize}
\item[1)] $\sigma\neq0,\;\;\; \omega\neq0$.

\item[2)] $\sigma\neq0,\;\;\; \omega= 0$.\label{dirac1}

\item[3)] $\sigma= 0, \;\;\;\omega\neq0$.\label{dirac2}

\item[4)] $\sigma= 0 = \omega, \;\;\;\mathbf{K}\neq0,\;\;\;
\mathbf{S}\neq0$.

\item[5)] $\sigma= 0 = \omega, \;\;\;\mathbf{K}= 0,
\;\;\;\mathbf{S}\neq0$.\label{elko1}

\item[6)] $\sigma= 0 = \omega, \;\;\; \mathbf{K}\neq0, \;\;\;
\mathbf{S} = 0$.
\end{itemize}

\noindent The current density $\mathbf{J}$ is always non-zero.
Types-(1), -(2), and -(3) spinor fields are denominated \textit{Dirac spinor
fields} for spin-1/2 particles and types-(4), -(5), and -(6) are
respectively called \textit{flag-dipole}, \textit{flagpole}\footnote{
Such spinor fields are constructed by a 
null 1-form field current and an also null 2-form field angular momentum, the ``flag'' \cite{koso}.} and
\textit{Weyl spinor fields}. Majorana spinor fields are a
particular case of a type-(5) spinor field. It is worthwhile to
point out a peculiar feature of types-(4), -(5) and -(6) spinor fields:
although $\mathbf{J}$ is always non-zero, 
$\mathbf{J}^{2}=-\mathbf{K}^{2}=0$. It shall be seen below that the
bilinear covariants related to an ELKO spinor field, satisfy
$\sigma=0=\omega,\;\;\mathbf{K}=0,\;\;\mathbf{S}\neq0$ and
$\mathbf{J}^{2}=0$.
Since Lounesto proved that there are \textit{no} other classes based on
distinctions among bilinear covariants, ELKO spinor fields
must belong to one of the disjoint six classes.

Types-(1), -(2) and -(3) Dirac spinor fields (DSFs) have different
algebraic and geometrical characters, and we would like to
emphasize the main differing points. For more details, see e.g. \cite{lou1,lou2}. 
Recall that if the quantities
$P = \sigma + {\bf J} + \g_{0123}\omega$ and $Q = {\bf S} + {\bf
K}\g_{0123}$ are defined \cite{lou1,lou2}, in type-(1) DSF we have
$P = -(\omega + \sigma\g_{0123})^{-1}{\bf K}Q$ and also $\psi =
-i(\omega + \sigma\g_{0123})^{-1}\psi$. In type-(2) DSF, $P$ is a
multiple of $\frac{1}{2\sigma} (\sigma + {\bf J})$ and looks like
a proper energy projection operator, commuting with the spin
projector operator given by $\frac{1}{2} (1 - i\g_{0123}{\bf
K}/\sigma)$. Also, $P = \g_{0123}{\bf K}Q/\sigma$. Further, in
type-(3) DSF, $P^2 = 0$ and $P = {\bf K}Q/\omega$.
The introduction of the spin-Clifford bundle makes it possible to consider all the
geometric and algebraic objects
---
 the Clifford bundle, spinor fields, differential form fields, operators and
Clifford fields --- as being elements of an
unique unified formalism. It is well known that spinor fields have three different,
although equivalent, definitions:
the operatorial, the classical and the algebraic one. In particular, the
operatorial definition allows us to factor --- up
to sign --- the DSF $\psi$ as $\psi = (\sigma + \omega\g_{0123})^{-1/2}R$, where $R \in$
Spin$^{e}_{1,3}$.
Denoting ${\bf K}_k = \psi\g_k\tilde\psi$, where $\tilde\psi$ denotes the
reversion of $\psi$, the set $\{{\bf J},{ K}_1,{ K}_2,{K}_3\}$ is
an orthogonal basis of $\RR^{1,3}$. On the other hand, in classes
(4), (5) and (6) --- where $\sigma = \bar\psi\psi = 0 = \omega =
\bar\psi\g_5\psi$, the vectors $\{{\bf J},{ K}_1,{ K}_2,{ K}_3\}$
no longer form a basis and collapse into a null-line \cite{lou1,lou2}. In such case
only the boundary term is non null. Finally, to a Weyl spinor 
field $\xi$ (type-(6)) with bilinear covariants {\bf J} and {\bf K}, 
 two Majorana spinor fields $\psi_\pm =
\frac{1}{2}(\xi + C(\xi))$ can be associated, where $C$ denotes the charge conjugation operator.
Penrose flagpoles are implicitly defined by the equation
$\sigma + {\bf J} + i{\bf S}- i\g_{0123}{\bf K}+ \g_{0123}\omega =
\frac{1}{2} ({\bf J} \mp i{\bf S}\g_{0123})$ \cite{lou1,lou2}. For a physically useful discussion regarding 
the disjoint classes -(5) and -(6) see, e.g., \cite{plaga}. 
The fact that two Majorana spinor fields $\psi_\pm$ can be written in terms of a Weyl type-(6) spinor field 
$\psi_\pm = \frac{1}{2}(\xi + C(\xi))$,
is an `accident' when the (Lorentzian) spacetime has $n= 4$ --- the present case --- or $n=6$ dimensions. The more general assertion 
concerns the property that two Majorana, and more generally ELKO
 spinor fields $\psi_\pm$ can be written in terms of a \emph{pure spinor} field --- hereon denoted 
by  $\mmu$ --- as $\psi_\pm = \frac{1}{2}(\mmu + C(\mmu))$. It is well known that Weyl spinor fields are pure spinor fields 
when $n=4$ and $n=6$. 
When the complexification of $\CC\otimes \RR^{1,3}$ of $\RR^{1,3}$ is considered, one can consider a maximal totally isotropic subspace $N$
  of $\CC^{1,3}$, by the Witt decomposition, where $\dim_\CC N = 2$. 
Pure spinors are defined by the property  $x\mmu = 0$ forall $x\in N \subset \CC^{1,3}$ \cite{cru}. 
In this context, Penrose flags can be defined by the expression Re$(i\mmu\tilde\mmu)$ \cite{benn}.

Now, the algebraic and formal
properties of ELKO spinor fields, as defined in \cite{allu,alu2,alu3,meu}, are briefly explored. An ELKO  $\Psi$
corresponding to a plane wave with momentum $p=(p^{0},\mathbf{p)}$
can be written, without loss of generality, as $\Psi(p)=\lambda({\bf p})
e^{-i{p\cdot x}}$ (or $\Psi(p)=\lambda({\bf p}) e^{i{p\cdot x}}$) where
\begin{equation}
\lambda({\bf p})=\binom{i\Theta\phi_{L}^{\ast}(\mathbf{p})}{\phi_{L}(\mathbf{p})},
\label{1}%
\end{equation}
\noindent and given the rotation generators denoted by
${\mathfrak{J}}$, the Wigner's spin-1/2 time reversal operator
$\Theta$ satisfies $\Theta
\mathfrak{J}\Theta^{-1}=-\mathfrak{J}^{\ast}$.
Hereon, as in \cite{allu}, the Weyl representation of
$\gamma^{\mu}$ is used, i.e.,
\begin{equation}
\gamma_{0}=\gamma^{0}=%
\begin{pmatrix}
\OO & \II\\
\II & \OO
\end{pmatrix}
,\quad-\gamma_{k}=\gamma^{k}=%
\begin{pmatrix}
\OO & -\sigma_{k}\\
\sigma_{k} & \OO
\end{pmatrix}
, \label{dirac matrices}%
\end{equation}
\noindent where
\begin{equation}
\II= \begin{pmatrix}
1 & 0\\
0 & 1
\end{pmatrix}
,\quad \OO=\begin{pmatrix}
0 & 0\\
0 & 0
\end{pmatrix}
,\quad
\sigma_{1}=%
\begin{pmatrix}
0 & 1\\
1 & 0
\end{pmatrix}
,\quad\sigma_{2}=%
\begin{pmatrix}
0 & -i\\
i & 0
\end{pmatrix}
,\quad\sigma_{3}=%
\begin{pmatrix}
1 & 0\\
0 & -1
\end{pmatrix},
\end{equation}
\noindent $\sigma_i$ are the Pauli matrices. Also, 
\begin{equation}
\gamma^{5}=i\gamma^{0}\gamma^{1}\gamma^{2}\gamma^{3}=i\gamma^{0123}=-i\gamma_{0123} = 
\begin{pmatrix}
\II&\OO \\
\OO & -\II
\end{pmatrix}.\end{equation}
\noindent  ELKO spinor fields are eigenspinors of the 
charge conjugation operator $C$, i.e., $C\lambda(\bf{p})=\pm \lambda({\bf
p})$, for
\begin{equation}
C=%
\begin{pmatrix}
\OO & i\Theta \\
-i\Theta & \OO
\end{pmatrix}
K .\label{conj}\end{equation} The operator $K$ is responsible for the $\mathbb{C}$-conjugation of Weyl 
spinor fields appearing on the right. The plus sign stands for {\it
self-conjugate} spinors, $\lambda^{S}({\bf p})$, while the minus
yields {\it anti self-conjugate} spinors, $\lambda^{A}({\bf p})$.
Explicitly, the complete form of ELKO spinor fields can be found by solving
the equation of helicity $(\sigma\cdot\widehat{\bf{p}})\phi^{\pm}=\pm \phi^{\pm}$ in the rest frame and
subsequently make a boost,  to recover the result for any ${\bf p}$
\cite{allu}. Here $\widehat{\bf{p}}:={\bf p}/\|{\bf p}\|$. The four spinor fields are given
\begin{equation}
\lambda^{S/A}_{\{\mp,\pm \}}({
p})=\sqrt{\frac{E+m}{2m}}\Bigg(1\mp
\frac{{\bf p}}{E+m}\Bigg)\lambda^{S/A}_{\{\mp,\pm \}}(\bf{0})
\label{form},\end{equation} where
\begin{equation}
\lambda_{\{\mp,\pm \}}(\bf{0})=%
\begin{pmatrix}
\pm i \Theta[\phi^{\pm}(\bf{0})]^{*} \\
\phi^{\pm}(\bf{0})
\end{pmatrix}
\label{four}.\end{equation} Note that, since
$\Theta[\phi^{\pm}(\bf{0})]^{*}$ and $\phi^{\pm}(\bf{0})$ have
opposite helicities, ELKO cannot be an eigenspinor field of the helicity
operator, and indeed carries both helicities. In
order to guarantee an invariant real norm, as well as positive
definite norm for two ELKO spinor fields, and negative definite norm for the
other two, the ELKO dual is given by
\begin{equation}
\overset{\neg}{\lambda}^{S/A}_{\{\mp,\pm \}}({\bf p})=\pm i \Big[
\lambda^{S/A}_{\{\pm,\mp \}}({\bf p})\Big]^{\dag}\gamma^{0}
\label{dual}.\end{equation}

Omitting the subindex of the spinor field $\phi_{L}(\mathbf{p})$, which
is denoted hereon by $\phi$, the left-handed spinor field
$\phi_{L}(\mathbf{p})$ can be represented by
\begin{equation}
\phi=\binom{\alpha(\mathbf{p})}{\beta(\mathbf{p})},\quad\alpha(\mathbf{p}%
),\beta(\mathbf{p})\in\mathbb{C}. \label{0}%
\end{equation}
\noindent\noindent Now using Eqs.(\ref{11}) it is possible to
\emph{calculate} explicitly the bilinear covariants for ELKO spinor fields\footnote{All the details are presented in 
 \cite{meu}.}:
\begin{align}
\mathring\sigma &  =\lambda^{\dagger}\gamma_{0}\lambda=0,\qquad
\mathring\omega   =-\lambda^{\dagger}\gamma_{0}\gamma_{0123}\lambda=0\label{16a}\\
\mathbf{\mathring J}  &  =\mathring J_{\mu}\gamma^{\mu}=\lambda^{\dagger}\gamma_{0}\gamma_{\mu}%
\lambda\gamma^{\mu}\neq 0\label{16}\\
\mathbf{\mathring K}  & =\mathring
K_{\mu}\gamma^{\mu}=\lambda^{\dagger}i\gamma_{123}\gamma_{\mu
}\lambda\gamma^{\mu}=0, \label{17}\\
\mathbf{\mathring S}  & =\frac{1}{2}\mathring
S_{\mu\nu}\gamma^{\mu\nu}=\frac{1}{2}\lambda^{\dagger
}\gamma_{0}i\gamma_{\mu\nu}\lambda\gamma^{\mu\nu}\neq 0
\label{18}%
\end{align}
\noindent From the formul\ae\, in Eqs.(\ref{16}, \ref{17}) it is
trivially seen that that $ \mathbf{J}\lrcorner\mathbf{K}=0.$
 Also, from Eq.(\ref{16}) it follows that
$ \mathbf{J}^{2}=0,$
 and it is immediate that all Fierz identities introduced by the
formul\ae\, in Eqs.(\ref{fi}) are trivially satisfied. 

 It is useful to
choose $i\Theta=\sigma_{2}$, as in \cite{allu}, in such a way that
it is possible to express
\begin{equation}
\lambda=\binom{\sigma_{2}\phi_{L}^{\ast}(\mathbf{p})}{\phi_{L}(\mathbf{p})}.
\label{01}%
\end{equation}

Now, any flagpole spinor field is an eigenspinor field of the charge
conjugation operator \cite{lou1,lou2}, which explicit action on a spinor $\psi$ is given by 
$\mathcal{C}\psi=-\gamma ^{2}\psi^{\ast}$. Indeed using Eq.(\ref{01}) it follows that
\begin{align}
-\gamma^{2}\lambda^{\ast}  & 
=\binom{\sigma_{2}\phi^{\ast}}{-\sigma_{2}\sigma_{2}^{\ast}\phi}\nonumber
=\lambda.
\end{align}
\noindent

Once the definition of ELKO spinor fields is recalled, 
we return to the previous discussion about Penrose flagpoles. Here we extend the definition of the Penrose 
poles, and we can prove that they are given in terms of an ELKO spinor field  by
the expression $\me\langle\lambda(\widetilde{\gamma_{0123}\lambda})\rangle_1$, and further, Penrose flags $F$ can also be written 
in terms of ELKO, as   $F = \me\langle\lambda(\widetilde{\gamma_{0123}\lambda})\rangle_2$. 
This assertion can be demonstrated following an  
reasoning analogous as the one exposed in \cite{benn,chev}.

\section{Which are the Dirac spinor fields that can be led to ELKO?}

In this Section we are interested in analyzing a matrix $M\in\CC(4)$ that
defines the transformation from an \emph{a priori} arbitrary DSF to an ELKO spinor field, i.e.,
\bege\label{cond} M\psi = \lambda.\enge\noi It shall be proved that not all DSFs can be led to ELKO, but only a subset 
of the three classes --- under Lounesto classification --- of DSFs restricted to some conditions. 
Explicitly we have
\begin{equation}
\begin{pmatrix}
M_{11} & M_{12} \\
M_{21} & M_{22}
\end{pmatrix}
\begin{pmatrix}
\phi_{R}({\bf p})\\
\phi_{L}({\bf p})
\end{pmatrix}
=%
\begin{pmatrix}
\epsilon \sigma_{2}\phi^{*}_{L}({\bf p})\\
\phi_{L}({\bf p})
\end{pmatrix}
\label{comeco},\end{equation} where $\epsilon=\pm 1$ and $M_{ij}\in\CC(2)$, ($i,j=1,2$). We are particularly interested
 to investigate the conditions imposed on DSFs
 that turn
them to be led to ELKO spinor fields. Taking into account that $\phi_{R}({\bf p})=\chi
\phi_{L}({\bf p})$, where $\chi = \frac{E + {\mathbf{\sigma}}\cdot
{\mathbf{p}}}{m}$ and $\ka\psi = \psi^*$ the following
system is obtained:
\begin{eqnarray}
M_{11}\chi + M_{12}=\epsilon \sigma_{2}\ka \nonumber
\\M_{21}\chi+M_{22}=1 \label{sitema}.\end{eqnarray} Then, writing
explicitly the entries of $M = [m_{pq}]_{p,q=1}^4$,  Eqs.(\ref{sitema}) read
\begin{eqnarray}
\chi m_{11}+m_{13}&=&0,\qquad\qquad \chi m_{31}+m_{33}=1,\nonumber\\
\chi m_{12}+m_{14}&=&-i\ka \epsilon, \quad\quad\, \chi m_{32}+m_{34}=0,\nonumber\\
\chi m_{21}+m_{23}&=&i\ka \epsilon, \qquad\quad \chi m_{41}+m_{43}=0,\nonumber\\
\chi m_{22}+m_{24}&=&0, \qquad\qquad \chi m_{42}+m_{44}=1,\nonumber\\
 \label{sistema}
\end{eqnarray}
in such way that the matrix $M$ can be written in the form
\begin{equation}
M=%
\begin{pmatrix}
m_{11} & m_{12} & -\chi m_{11} & -i\epsilon \ka -\chi m_{12}\\
m_{21} & m_{22} & i\epsilon \ka-\chi m_{21} & -\chi m_{22}\\
m_{31} & m_{32} & 1-\chi m_{31} & -\chi m_{32}\\
m_{41} & m_{42} & -\chi m_{41} & 1- \chi m_{42}\\
\end{pmatrix}
\label{eme2}.\end{equation} In order to have the product
$(-i\epsilon \ka -m_{12})(i\epsilon \ka-\chi m_{21})$ equal to
$(i\epsilon \ka-\chi m_{21})(-i\epsilon \ka -m_{12})$, which can
be useful in further calculations, we take $m_{12}=-m_{21}$. From
now on, in order to completely fix the matrix $M$, the {\it ansatz}
\begin{eqnarray}
m_{11}=m_{22}=0= m_{32}=m_{41},\nonumber\\
m_{31}=m_{42}=1=m_{12} \label{fix}
\end{eqnarray} is regarded, and  $M$ is written as
\begin{equation}
M=%
\begin{pmatrix}
0 & 1 & 0 & -i\epsilon \ka -\chi \\
-1 & 0 & i\epsilon \ka+\chi & 0\\
1 & 0 & 1-\chi & 0\\
0 & 1 & 0 & 1- \chi \\
\end{pmatrix}
\label{ansatz}.\end{equation}
\noindent
Note that such matrix is not unitary, and since $\det M\neq 0$, there exists (see Eq.(\ref{cond})) $M^{-1}$ such that
$\psi=M^{-1}\lambda$. Besides, it is immediate to note that \bege
\bar{\psi}:=\psi^\dagger\gamma^0=\lambda^{\dag}(M^{-1})^{\dag}\gamma^{0},\label{j1}
\enge such that $\bar\psi$ can be related to the ELKO dual by
\bege\bar{\psi} =\mp
i\overset{\neg}{\lambda}^{S/A}_{\{\mp,\pm\}}\gamma^{0}(M^{-1})^{\dag}\gamma^{0}.\label{j2}
\enge In what follows, the matrix $M$ establishes necessary
conditions on the Dirac spinor fields under which the mapping
given by Eq.(\ref{cond}) is satisfied. However, the {\it ansatz}
in Eq.(\ref{ansatz}) has just an illustrative r\^ole. In fact, for any
matrix satisfying Eq.(\ref{eme2}), there are corresponding constraints on the
components of DSFs. Hereafter, we shall calculate the
conditions to the case where {\bf p} = 0 (and
consequently $\chi = 1$), since a Lorentz boost can be implemented on the rest frame in the constraints.
Anyway, without lost of generality, the conditions to be found on DSFs must hold in all referentials, 
and in particular in the rest frame corresponding to {\bf p} = 0.

Substituting Eq.(\ref{cond}) in the definition given by
Eqs.(\ref{fierz}) we have
\begin{align}
{}\mathring\sigma &  =\psi^\dagger M^\dagger\gamma_{0}M\psi,\quad
{\mathbf{\mathring J}}=\mathring{J}_{\mu}\gamma^{\mu
}=\psi^{\dagger}M^\dagger\gamma_{0}\gamma_{\mu}M\psi\gamma^{\mu},\quad\mathbf{\mathring S}
=\mathring{S}_{\mu\nu}\gamma^{\mu\nu}=\frac{1}{2}\psi^{\dagger}M^\dagger\gamma_{0}i\gamma_{\mu\nu
}M\psi\gamma^{\mu\nu},\nonumber\\
\quad\mathbf{\mathring K}  &  = \mathring{K}_\mu\gamma^\mu =
i\psi^{\dagger}M^\dagger\gamma_{0}\gamma_{0123}\gamma_{\mu}%
M\psi\gamma^{\mu},\quad\mathring\omega=-\psi^{\dagger}M^\dagger\gamma_{0}\gamma_{0123}M\psi.
\label{111}
\end{align}  These new bilinear covariants --- expressed in terms of DSFs --- 
 are related to ELKO spinor fields, and by the definition of type-(5) spinor fields under Lounesto classification, 
they  automatically satisfy
the conditions $\mathring\sigma= 0 = \mathring\omega, 
\;\mathbf{\mathring K}= 0,$ and $\mathbf{\mathring S}\neq0$. Types-(1), -(2), and -(3) of 
DSFs satisfy {\bf K}$\neq 0$, but when they are
transformed in ELKO spinor fields via the action of $M$, they must
satisfy $\mathbf{\mathring{K}}= \mathring{K}^\mu\g_\mu = 0$. As $\{\g_\mu\}$ is a basis of
$\RR^{1,3}$, each one of the components $\mathring{K}^\mu$ \emph{must equal zero},
i.e.,
\beq \mathring{K}_0 &=&
  \psi^{\dagger}M^\dagger\gamma_{0}i\gamma_{0123}\gamma_0M\psi\n &=&
 \psi^{\dagger}\Bigg[ \begin{pmatrix}0&a\\ a^{*}&0\\
\end{pmatrix} \otimes \II \Bigg] \psi\n
&=&0\label{negrini10}\eeq \noi where $a:=-(1+i\epsilon \ka)$. The
other components read
\beq \mathring{K}_1 &=&
\psi^{\dagger}M^\dagger\gamma_{0}i\gamma_{0123}\gamma_1M\psi\n
&=& -  \psi^{\dagger}\Bigg[ \begin{pmatrix}0&a\\
a^{*}&0\\
\end{pmatrix}\otimes \sigma_{1} \Bigg] \psi\n &=&0\label{negrini}\\
\mathring{K}_2 &=&
\psi^{\dagger}M^\dagger\gamma_{0}i\gamma_{0123}\gamma_2M\psi\n
&=&   \psi^{\dagger}\Bigg[ \begin{pmatrix}2 & a\\
a&-a\\
\end{pmatrix} \otimes \sigma_{2} \Bigg] \psi\n &=&0\label{negrini3}\\
\mathring{K}_3 &=&
\psi^{\dagger}M^\dagger\gamma_{0}i\gamma_{0123}\gamma_3M\psi\n
&=&   \psi^{\dagger}\Bigg[ \begin{pmatrix}0&-a \\
a^{*}&0\\
\end{pmatrix} \otimes \sigma_{3} \Bigg]  \psi\n &=&0\label{negrini5}.\eeq \noi
After all, denoting $\psi = (\psi_1,\psi_2,\psi_3,\psi_4)^T$ ($\psi_r \in \CC, r=1,\ldots,4$),
 we have the simultaneous conditions for
Eqs.(\ref{negrini10})-(\ref{negrini5}) respectively:
\begin{align}
0&= \mathbb{R}{\rm e}(\psi_1^*\psi_3)+\mathbb{R}{\rm e}(\psi_2^*\psi_4)\n 0&=
\mathbb{R}{\rm e} (\psi_2^*\psi_3)+ \mathbb{R}{\rm e} (\psi_1^*\psi_4)\n 0&=
\mathrm{Im}(\psi_1^*\psi_4)-\mathrm{Im}(\psi_2^*\psi_3)-2\mathrm{Im}(\psi_3^*\psi_4)-2\mathrm{Im}(\psi_1^*\psi_2)
\n 0&= \mathbb{R}{\rm e}(\psi_1^*\psi_3)-\mathbb{R}{\rm e}(\psi_2^*\psi_4)
\label{partes}.\end{align} These constraints must hold for
types-(1), -(2), and -(3) DSFs. Note that the first and the
last conditions together mean $\mathbb{R}{\rm e}(\psi_1^*\psi_3)=0$ and
$\mathbb{R}{\rm e}(\psi_2^*\psi_4)=0$. In what follows we obtain the
extra necessary and sufficient conditions for each class of DSFs.

\subsection{Additional conditions on class-(2) Dirac spinor fields}
\label{sub2}Type-(2) DSFs satisfy  by definition the condition
\beq \omega&=& -\psi^\dagger\gamma_0\gamma_{0123}\psi\n &=&
-\psi_1^*\psi_3 - \psi_3^*\psi_1 + \psi_2^*\psi_4 + \psi_4^*\psi_2 =
0 .\eeq Besides, the conditions obtained from $\mathring{\bf  K}=0$, we also
have in this case the additional condition: \beq \mathring\sigma
&=&\psi^\dagger M^\dagger\gamma_{0}M\psi\n
&=& \psi^\dagger \Bigg[\begin{pmatrix}0&a\\
-a^{*}&0\\ \end{pmatrix} \otimes i\sigma_{2} \Bigg ]\psi\n &=&
\mathbb{R}{\rm e}(\psi_1^*\psi_4)+\mathrm{Im}(\psi_2^*\psi_3)\n
&=&0\label{ad2}.\eeq

\subsection{Additional conditions on class-(3) Dirac spinor fields}
\label{sub3}Class-(3) Dirac spinor fields satisfy --- by definition --- the condition
\beq \sigma&=& \psi^\dagger\gamma_0\psi\n &=&
|\psi_1|^2+|\psi_2|^2-|\psi_3|^2-|\psi_4|^2=0 .\eeq Apart of the
conditions obtained from $\mathring{\bf K}=0$, we also have for this class
the additional condition: \beq \mathring\omega &=&-\psi^\dagger
M^\dagger\gamma_{0}\gamma_{0123} M\psi\n
&=& \psi^\dagger \Bigg[\begin{pmatrix}2&a\\
a&0\\ \end{pmatrix} \otimes \sigma_{2} \Bigg]\psi\n &=&
\mathrm{Im}(\psi_1^*\psi_4)-\mathrm{Im}(\psi_2^*\psi_3)-2\mathrm{Im}(\psi_1^*\psi_2)\n
&=&0\label{ad3}.\eeq

\subsection{Additional conditions on class-(1) Dirac spinor fields}
\label{sub1} After the action of  the matrix $M$, class-(1) DSFs must
obey all the conditions given by Eqs.(\ref{partes}), (\ref{ad2}),
and (\ref{ad3}). Note that if one relaxes the condition given by Eq.(\ref{ad2}) or
Eq.(\ref{ad3}), DSFs of types-(3) and -(2) are respectively obtained.

Using the decomposition $\psi_j=\psi_{ja}+i\psi_{jb}$ (where $\psi_{ja}$ = $\mathbb{R}$e($\psi_j$) and $\psi_{jb}$ = Im($\psi_j$)) it follows that
$\mathbb{R}{\rm e}(\psi_i^*\psi_j)=\psi_{ia}\psi_{ja}+\psi_{ib}\psi_{jb}$
and
$\mathrm{Im}(\psi_i^*\psi_j)=\psi_{ia}\psi_{jb}-\psi_{ib}\psi_{ja}$
for $i,j=1,\ldots,4$. So, in components, the conditions in common for
all types of DSFs are \ba
\psi_{1a}\psi_{3a}+\psi_{1b}\psi_{3b}&=&0 \label{c1},\\
\psi_{2a}\psi_{4a}+\psi_{2b}\psi_{4b}&=&0 \label{c2}, \ea and the
additional conditions for each case are summarized in Table I below.
\begin{center}
\begin{table}[!h]
\begin{tabular}{|c|c|}
  \hline
  {\bf Class} & {\bf Additional conditions}  \\
  \hline
 \hline (1) & $\psi_{2a}(\psi_{3a}-\psi_{3b})+\psi_{2b}(\psi_{3a}+\psi_{3b}) = 0 = \psi_{3a}\psi_{4b}-\psi_{3b}\psi_{4a}$ \\\hline
  (2) & $\psi_{3a}\psi_{4b}-\psi_{3b}\psi_{4a} = 0 = \psi_{2a}\psi_{3a}+\psi_{2b}\psi_{3b}+\psi_{1a}\psi_{4a}+\psi_{1b}\psi_{4b}$
\\\hline
  (3) & $\psi_{2a}(\psi_{3a}-\psi_{3b})+\psi_{2b}(\psi_{3a}+\psi_{3b})=0$
 and \\
{}&$(\psi_{1a}\psi_{4b}-\psi_{1b}\psi_{4a})-(\psi_{2a}\psi_{3b}-\psi_{2b}\psi_{3a})-2(\psi_{3a}\psi_{4b}-\psi_{3b}\psi_{4a})-$
 $2(\psi_{1a}\psi_{2b}-\psi_{1b}\psi_{2a})=0$ \\
  \hline
\end{tabular}
\caption{Additional conditions, in components, for class (1), (2)
and (3) Dirac spinor fields.}
\end{table}
\end{center}

\section{concluding remarks and outlooks}
Once the matrix $M$ --- leading an arbitrary DSF to an ELKO ---  has been introduced, we proved that it can be written
 in the general form given by Eq.(\ref{eme2}), without loss of generality. The \emph{ansatz} given by Eq.(\ref{eme2})
is useful to illustrate and explicitly exhibit how to obtain the necessary conditions on the components 
of a DSF --- under Lounesto spinor field classification --- in order to it be led to an ELKO spinor field.
In the case of a type-(1) DSF, as accomplished in Subsec.(\ref{sub1}), there are six conditions, from the definition of ELKO
($\mr{\sigma}= 0 = \mr{\omega} = \mr{K}^\mu)$, and then the equivalence class of type-(1) DSFs
 that can be led to ELKO spinor fields  can be written in the form\footnote{Among the
 three equivalent definitions of spinor fields, viz., the classical, algebraic, and operatorial, here 
the classical one --- where a spinor is an element that carries the representation space of the group Spin$_+$(1,3),
 is regarded.}
\bege
\psi=\begin{pmatrix}
\psi_1\\
f_1(\psi_1)\\
f_2(\psi_1)\\
f_3(\psi_1)\\
\end{pmatrix}
\enge\noi where $f_i$ are complex scalar functions of the component $\psi_1\in\mathbb{C}$ of $\psi$, obtainable --- using the implicit function theorem --- 
through the conditions given in Eqs.(\ref{c1}), (\ref{c2}), and also those 
given by Table I.
For a general and arbitrary \emph{ansatz}, the equivalence class of type-(1) DSFs
 that can be led to ELKO spinor fields, via the matrix $M$, are given by 
\bege\psi=\begin{pmatrix}
\psi_1\\
g_1(M)(\psi_1)\\
g_2(M)(\psi_1)\\
g_3(M)(\psi_1)\\
\end{pmatrix}
\enge\noi where each $g_i(M)$ is a complex scalar function of the component $\psi_1\in\mathbb{C}$ of $\psi$.
Such scalar functions depend explicitly on the form of $M$, and to a fixed but arbitrary $M$ there corresponds other six conditions analogous to Eqs.(\ref{c1}), (\ref{c2}), and also those 
given by Table I. All these conditions 
obtained by the \emph{ansatz} is general, and illustrates the general procedure of finding the conditions.

Regarding Subsecs. (\ref{sub2}) and (\ref{sub3}), 
for the equivalence class of type-(2) and -(3) DSFs that are led to ELKO spinor fields,
it is only demanded five conditions, instead of six, since respectively $\mathring\sigma=0, \mathring\omega\neq 0$, and 
$\mathring\sigma\neq 0, \mathring\omega = 0$. In both cases, the most general form of the DSFs are given by
\bege\psi=
\begin{pmatrix}
\psi_{1a} + i\psi_{1b}\\
\psi_{2a} + i\psi_{2b}\\
\psi_{3a} + i\psi_{3b}\\
\psi_{4a} + i\psi_{4b}\\
\end{pmatrix}=\begin{pmatrix}
\psi_{1a} + i\psi_{1b}\\
\psi_{2a} + i h_1(M)(\psi_{1a},\psi_{1b},\psi_{2a})\\
h_2(M)(\psi_{1a},\psi_{1b},\psi_{2a}) + i h_3(M)(\psi_{1a},\psi_{1b},\psi_{2a})\\
h_4(M)(\psi_{1a},\psi_{1b},\psi_{2a}) + i h_5(M)(\psi_{1a},\psi_{1b},\psi_{2a})\\
\end{pmatrix}
\enge\noi  where each $h_A(M)$ ($A=1,\ldots,5$) is a $M$ matrix-dependent 
real scalar function of the (real) components $\psi_{1a},\psi_{1b},\psi_{2a}$ 
 of $\psi$.

One of the main physical motivations here  is that 
dark matter, which can be described by ELKO \cite{alu3},
interacts very weakly with  Standard Model (SM) particles, and the task is how to extend SM in order to incorporate
ELKO. This
approach can be of prime importance in a posteriori
investigation about the dynamical aspects and about the Standard Model in
ELKO context. Once we know the behaviour of DSFs in the context of SM, and also the particular subsets of 
the equivalence classes of DSFs that can be
led to ELKO, it is natural to ask whether it is now possible to extend SM using ELKO. Our paper 
is the first attempt --- up to our knowledge ---
to accomplish this purpose, and a new and physically alluring branch on Standard Model extensions and cosmology 
is proposed for further promising investigations.



\section{Acknowledgment}
The authors are very grateful to Prof. Dharamvir Ahluwalia-Khalilova for important comments
about this paper, and to JMP referee for pointing out elucidating and enlightening viewpoints.
Rold\~ao da Rocha thanks to Funda\c c\~ao de Amparo \`a Pesquisa
do Estado de S\~ao Paulo (FAPESP) (PDJ 05/03071-0) and J. M. Hoff
da Silva thanks to CAPES-Brazil for financial support.


\begin{thebibliography}{99}


 \bibitem {allu}D. V. Ahluwalia-Khalilova and D. Grumiller, \emph{Spin Half
Fermions, with Mass Dimension One: Theory, Phenomenology, and Dark Matter},
\textit{JCAP} \textbf{07} (2005) 012 [\texttt{arXiv:hep-th/0412080v3}].

\bibitem{alu2} D. V. Ahluwalia-Khalilova and D. Grumiller, 
\emph{Dark matter: A spin one half fermion field with mass dimension one?},  
    Phys. Rev. {\bf D72} (2005) 067701  [{\tt arXiv:hep-th/0410192v2}]. 

\bibitem{alu4} D. V. Ahluwalia-Khalilova, 
\emph{Extended set of Majorana spinors, a new dispersion relation, and a preferred frame},
[{\tt arXiv:hep-ph/0305336v1}].


\bibitem{mac} J. A. McLennan, \emph{Parity nonconservation and the theory of neutrino}, Phys. Rev. {\bf 106}
(1957) 821-822.

\bibitem{case} K. M. Case, \emph{Reformulation of Majorana theory of the neutrino}, Phys. Rev. {\bf 107} (1957) 307-316.

\bibitem{alu3} Ahluwalia-Khalilova D V, 
\emph{Dark matter, and its darkness},  
    Int. J. Mod. Phys. {\bf D15} (2006) 2267-2278  [{\tt arxiv:hep-th/0603545v3}]. 

\bibitem{ani} A. Anisimov, \emph{Majorana dark matter}, 
 ``6th International Workshop on the Identification of Dark Matter'', Rhodes, Greece, Sept 2006 
[{\tt arXiv:hep-ph/0612024v2}].

\bibitem{123} P. Frampton, S. Glashow, and T. Yanagida, \emph{Neutrinoless double beta decay can constrain neutrino dark matter}, 
     Phys. Lett. {\bf B532} (2002) 15-18 [{\tt arXiv:hep-ph/0201262}].



\bibitem{jee} M. J. Jee et al, \emph{Discovery of a Ringlike Dark Matter Structure in the Core of 
the Galaxy Cluster Cl 0024+17}, Astrophys. J. {\bf 661} (2007) 728-749 [{\tt arXiv:0705.2171v1}]




\bibitem {moro}{R. A. Mosna and W. A. Rodrigues, Jr., \emph{The bundles of
algebraic and Dirac-Hestenes spinor fields}, J. Math. Phys.
\textbf{45} (2004) 2945-2988 [{\tt
arXiv:math-ph/0212033v5}]


\bibitem {rod}W. A. Rodrigues, Jr., \emph{Algebraic and Dirac Hestenes Spinors and
Spinor Fields}, {J. Math. Phys}. \textbf{45} (2004) 2908-2966
[\texttt{arXiv:math-ph/0212030v6.}}]


\bibitem{op} V. Figueiredo, E. Capelas de Oliveira, W. A. Rodrigues, Jr., 
\emph{Covariant, algebraic, and operator spinors}, Int. J. Theor. Phys. {\bf 29} (1990) 371-395.


\bibitem{hes} D. Hestenes, \emph{Real Spinor Fields}, J. Math. Phys. {\bf 8} (1967) 798-808.

\bibitem{hes1} D. Hestenes and G. Sobczyk, \emph{Clifford Algebra to Geometric Calculus: 
A Unified Language for Mathematics and Physics}, D. Reidel, Dordrecht 1984.

\bibitem{cru} A. Crumeyrolle, \emph{Orthogonal and Symplectic Clifford Algebras: 
Spinor Structures, Kluwer Academic}, Dordrecht 1990.

\bibitem{benn} I. M. Benn and R. W. Tucker, \emph{
An Introduction to Spinors and Geometry with Applications in Physics}, Adam Hilger, Bristol 1987.

\bibitem{cartan} E. Cartan, \emph{The Theory of Spinors}, translated from \emph{Leçons sur la théorie des spineurs}, 
1937, Dover, New York 1966.

\bibitem{chev} C. Chevalley, \emph{The Algebraic Theory of Spinors}, Columbia University Press, New York 1954.

\bibitem{riesz} M. Riesz, \emph{Clifford Numbers and Spinors}, University of Maryland Press, College Park 1958.

\bibitem {lou1}P. Lounesto, \emph{Clifford Algebras, Relativity and Quantum
Mechanics}, in Letelier P and Rodrigues W A, Jr. (eds.),
\emph{Gravitation: the Spacetime Structure}, Proc. of the
$8^{\mathrm{th}}$ Latin American Symposium on Relativity and
Gravitation, \'{A}guas de Lind\'{o}ia, Brazil, 25-30 July 1993,
World-Scientific, London 1993.


\bibitem{lou2}P. Lounesto, \emph{Clifford Algebras and Spinors},
2$^{\mathrm{nd}}$ ed., pp. 152-173, Cambridge Univ. Press,
Cambridge 2002.


\bibitem{meu} R. da Rocha and  W. A. Rodrigues, Jr.,
\emph{Where are ELKO spinors in Lounesto spinor field
classification?},
 Mod. Phys. Lett. {\bf A21} (2006) 65-74 [{\tt arXiv:math-ph/0506075v3}].


\bibitem{boe1} C. G. Boehmer, \emph{The Einstein-Elko system -- Can dark matter drive inflation?}, 
[{\tt arXiv:gr-qc/0701087v1}]

\bibitem{boe2}  C. G. Boehmer, \emph{The Einstein-Cartan-Elko system}, 
Annalen Phys. {\bf 16} (2007) 38-44 [{\tt arxiv:gr-qc/0607088v1}]. 




\bibitem{hor} S. Holst,  \emph{Barbero's Hamiltonian derived from a generalized Hilbert-Palatini action}, 
Phys. Rev. {\bf D53} (1996) 5966-5969 [{\tt arXiv:gr-qc/9511026v1}].

\bibitem{tn1} J. M. Nester  and R. S. Tung,  
    \emph{A Quadratic Spinor Lagrangian for General Relativity}, Gen. Rel. Grav. {\bf 27} (1995) 115-119 
[{\tt arXiv:gr-qc/9407004v1}].


\bibitem{bars1} I. Bars and S. W. MacDowell, \emph{A spin-3/2 theory of gravitation},  Gen. Rel. Grav. {\bf 10} (1979)
205-209.



\bibitem{jpe} R. da Rocha and J. G. Pereira, 
\emph{The quadratic spinor Lagrangian, axial torsion current, and generalizations}, to appear in Int. J. Mod. Phys. {\bf D}
[{\tt arXiv:gr-qc/0703076v1}].


\bibitem{ahlu4} D. V. Ahluwalia-Khalilova, \emph{Theory of neutral particles: Mclennan-Case construct for neutrino, its
generalization, and a new wave equation}, Int. J. Mod. Phys. {\bf A11} (1996) 1855-1874 [{\tt arXiv:hep-th/9409134v2}].    



\bibitem{pe1} R. Penrose, \emph{Twistor algebra}, J. Math. Phys. {\bf 8} (1967) 345-366.

\bibitem{pe2} R. Penrose and W. Rindler, 
\emph{Spinors and Spacetime vol.2: Spinor and Twistor Methods in Spacetime Geometry}, Cambridge Univ. Press, Cambridge 1986.

\bibitem{pe3} R. Penrose, \emph{Twistor quantisation and curved space-time}, 
Int. J. Theor. Phys. {\bf 1}  (1968) 61-99.

\bibitem{pos1} R. da Rocha and  J. Vaz, Jr., \emph{Twistors, Generalizations and Exceptional Structures},
 PoS WC2004 (2004) 022 [{\tt arXiv:math-ph/0412037v2}].

\bibitem{rocha1}  R. da Rocha and  J. Vaz, Jr., 
\emph{Revisiting Clifford algebras and spinors II: Weyl spinors in Cl(3,0) and Cl(0,3) and the Dirac equation}, 
[{\tt arXiv:math-ph/0412075v1}]. 

\bibitem{dor2} C. Doran, A. Lasenby A, and S. Gull, 
\emph{States and operators in the spacetime algebra}, Found. Phys. {\bf 23} (1993) 1239-1264.

\bibitem{bette} A. Bette, \emph{Twistors, special relativity, conformal symmetry and minimal coupling - a review}, 
Int. J. Geom. Meth. Mod. Phys. {\bf 2} (2005) 265-304
[{\tt arXiv:hep-th/0402150v2}].
   
\bibitem{bette2} A. Bette, \emph{Twistor dynamics of a massless spinning particle}, {\bf 40} (2001)
377-386.

\bibitem{espo} G. Esposito, \emph{From spinor geometry to complex general relativity}, Int. J. Geom. Meth. Mod. Phys. {\bf 2} (2005) 675-731,
[{\tt arXiv:hep-th/0504089v2}].

\bibitem{ro4} W. A. Rodrigues, Jr., R. da Rocha, and J. Vaz, Jr.,
\emph{Hidden Consequence of Active Local Lorentz Invariance},
Int. J. Geom. Meth. Mod. Phys.  {\bf 2} (2005)
 305-357 [{\tt arXiv:math-ph/0501064v6}].

\bibitem{wal} W. A. Rodrigues, Jr. and E. Capelas de Oliveira, \emph{The Many Faces of Maxwell, Dirac 
and Einstein Equations. A Clifford Bundle Approach}, Lecture Notes in Physics {\bf 722}, Springer, New York 2007. 

\bibitem {lawmi}{H. B. Lawson, Jr. and M. L. Michelson,
\textit{Spin Geometry}, Princeton University Press, Princeton 1989.}


\bibitem {choquet}{Y. Choquet-Bruhat, C. DeWitt-Morette, and
M. Dillard-Bleick, \emph{Analysis, Manifolds and Physics (revised
edition)}, North-Holland Publ. Co, Amsterdam 1977.}

\bibitem {cra}J. P. Crawford, \emph{On the Algebra of Dirac Bispinor Densities:
Factorization and Inversion Theorems}, J. Math. Phys. \textbf{26}
(1985) 1429-1441.




\bibitem{holl}P. R. Holland, \emph{Relativistic Algebraic Spinors and Quantum
Motions in Phase Space}, {Found. Phys.} \textbf{16} (1986)
708-709.

\bibitem{hol}P. R. Holland, \emph{ Minimal Ideals and Clifford
Algebras in the Phase Space Representation of spin-1/2 Fields}, p.
273-283 in Chisholm J S R and Common A K (eds.), \emph{Proceedings
of the Workshop on Clifford Algebras and their Applications in
Mathematical Physics (Canterbury 1985)}, Reidel, Dordrecht 1986.


\bibitem{koso} 
M. R. Francis and A. Kosowsky, \emph{The construction of spinors in geometric algebra}, 
Annals Phys. {\bf 317} (2005) 383-409 [{\tt arXiv:math-ph/0403040v2}]. 












\bibitem {qtmosna} R. A. Mosna and J. Vaz Jr., \emph{Quantum Tomography for Dirac
Spinors}, {Phys. Lett.} \textbf{A315} (2003) 418-425
[\texttt{arXiv:quant-ph/0303072v2}].


\bibitem {plaga}R. Plaga, \emph{A Demonstration that the Observed Neutrinos are not
Majorana Particles},  \ [\texttt{arXiv:hep-ph/9610545v3}];
\emph{The non-equivalence of Weyl and Majorana neutrinos with
standard-model gauge interactions},
[\texttt{arXiv:hep-ph/0108052v1}].








\end{thebibliography}
\end{document}